\documentclass[aps,amsfont,amssymb,twocolumn,floats,pre,showpacs,superscriptaddress,floatfix]{revtex4}
\usepackage{graphicx}
\newcommand{\sect}[1]{\setcounter{equation}{0}\section{#1}}

\newcommand{\bfm}[1]{\mbox{\boldmath${#1}$}}
\begin{document}
\title{Two-parameter deformations of logarithm, exponential,
and entropy: A consistent framework for generalized statistical
mechanics}
\author{G. Kaniadakis$^1$}\email{giorgio.kaniadakis@polito.it}
\author{M. Lissia$^2$}\email{marcello.lissia@ca.infn.it}
\author{A.M. Scarfone$^{1,2}$}\email{antonio.scarfone@polito.it}
\affiliation{$^1$Dipartimento di Fisica and Istituto Nazionale di
Fisica della Materia (INFM),\\ Politecnico di Torino, Corso Duca
degli Abruzzi 24, 10129 Torino, Italy} \affiliation{$^2$ Istituto
Nazionale di Fisica Nucleare (INFN) and Physics Department,
Universit\`a di Cagliari, I-09042 Monserrato (CA), Italy}
\begin {abstract}
A consistent generalization of statistical mechanics is obtained
by applying the maximum entropy principle to a trace-form entropy
and by requiring that physically motivated mathematical properties
are preserved. The emerging differential-functional equation
yields a two-parameter class of generalized logarithms, from which
entropies and power-law distributions follow: these distributions
could be relevant in many anomalous systems. Within the specified
range of parameters, these entropies possess positivity,
continuity, symmetry, expansibility, decisivity, maximality,
concavity, and are Lesche stable. The Boltzmann-Shannon entropy
and some one parameter generalized entropies already known belong
to this class. These entropies and their distribution functions
are compared, and the corresponding deformed algebras are
discussed.
\end {abstract}
\date{24 September 2004; revised 16 February 2005}
\pacs{02.50.-r, 05.20.-y, 05.90.+m}

\maketitle

\sect{Introduction}
\label{sec:intro}

In recent years the study of an increasing number of natural
phenomena that appear to deviate from standard statistical
distributions has kindled interest in alternative formulations of
statistical mechanics.

Among the large class of phenomena which show asymptotic power-law
behaviors, we recall anomalous diffusion
\cite{Bouchaud,Shlesinger2}, turbulence \cite{Boghosian,Beck1},
transverse-momentum distribution of hadron jets in e$^+$e$^-$
collisions \cite{Bediaga}, thermalization of heavy quarks in
collisional process \cite{Walton}, astrophysics with long-range
interaction \cite{Binney} and others
\cite{Kaniadakis0,Kaniadakis00}. Typically, anomalous systems have
 multifractal or hierarchical structure, long-time memory, and
long-range interaction \cite{Landsberg2,Hanel}.

The success of the Boltzmann-Gibbs (BG) theory has suggested that
new formulations of statistical mechanics should preserve most of
the mathematical and epistemological structure of the classical
theory, while reproducing the emerging phenomenology of anomalous
systems. To this end, new entropic forms have been introduced,
which would generalized the classical one introduced by Boltzmann
and Gibbs and, successively, by Shannon in a different context
(BGS entropy)
\begin{eqnarray}
S_{_{\rm BGS}}=-\sum_{i=1}^Np_{i}\,\ln p_{i} \ .\label{BGSentropy}
\end{eqnarray}

There is no systematic way of deriving the ``right'' entropy for a
given dynamical system. Among the many generalizations of the BGS
entropy, one can find the entropies considered by R\'enyi
\cite{Renyi}, by Tsallis ($q$ entropy) \cite{Tsallis3}, by Abe
\cite{Abe3}, by Tsallis, Mendes and Plastino (escort entropy)
\cite{Tsallis4}, by Landsberg and Vedral \cite{Landsberg}, and
recently  by Kaniadakis (${\kappa}$ entropy)
\cite{Kaniadakis1,Kaniadakis2}. For a historical outline see Ref.
\cite{editorial}.

Generalizations lead to abandoning part of the original
mathematical structure and properties. For instance, it is known
that the BGS entropy is of trace form \cite{Scarfone2,Csiszar}
\begin{eqnarray}
S=-\sum_{i=1}^Np_{i}\,\Lambda(p_{i})=-\langle
\Lambda(p_i)\rangle,\label{ggent}
\end{eqnarray}
with $\Lambda(p_i)=\ln(p_i)$; on the contrary the R\'enyi entropy,
the Landsberg-Vedral entropy, and the escort entropy are not of
trace form. The R\'enyi entropy and the Landsberg-Vedral entropy
are concave only for $0<q<1$, while the escort entropy is concave
only for $q>1$ \cite{Tsallis5}.

A  fundamental test for a statistical functional ${\cal O}(p)$ of
the probability distribution to be physically meaningful is given
by the Lesche stability condition \cite{Scarfone}: the relative
variation $\Big|[{\cal O}(p)-{\cal O}(p^\prime)]/{\rm sup}[{\cal
O}(p)]\Big|$ should go to zero in the limit that the probabilities
$p_{_i}\rightarrow p_{_i}^\prime$. This stability condition for a
functional is a necessary but not sufficient condition for the
existence of an associated observable. Lesche \cite{Lesche} showed
that, adopting the measure
$|\!|p-p^\prime|\!|_1=\sum_i|p_{_i}-p_{_i}^\prime|$ as estimator
of the closure of the two distributions, the BGS entropy is
stable, while the R\'enyi entropy is unstable (except for the
limiting case $q=1$ corresponding to the BGS entropy)
\cite{Arimitsu,Lesche1,Yamano}. Also the Landsberg-Vedral entropy
and the escort entropy do not satisfy the Lesche criterion
\cite{Tsallis5}. On the other hand it is already known that  the
Abe entropy \cite{Scarfone}, the q entropy \cite{Abe4}, and the
$\kappa$ entropy \cite{Scarfone1} are stable.

In the present paper, a natural continuation of the work in
Ref.~\cite{Scarfone2}, we consider the trace-form entropy given by
Eq. (\ref{ggent}), where $\Lambda(x)$ is an arbitrary analytic
function that represents a generalized version of the logarithm,
while its inverse function is the corresponding generalized
exponential \cite{Naudts1,Naudts2,Naudts3,Naudts4}. A consistent
framework is maintained with the use of the maximum entropy
(MaxEnt) principle. This approach yields a two-parameter class of
nonstandard entropies introduced a quarter of a century ago by
Mittal \cite{Mittal} and Sharma, Taneja \cite{Taneja1} and
successively studied by Borges and Roditi in Ref.~\cite{Borges1}.
It automatically unifies the entropies introduced by Tsallis
~\cite{Tsallis3}, Abe ~\cite{Abe3}, and Kaniadakis
~\cite{Kaniadakis1,Kaniadakis2}.

The paper is organized as follows. In Sec.~\ref{sec:canonical} the
differential-functional equation for the deformed logarithm
$\Lambda(x)$ proposed in ~\cite{Kaniadakis2} is briefly
reconsidered within the canonical ensemble formalism. In
Sec.~\ref{sec:solution} we solve this equation obtaining the more
general explicit form of $\Lambda(x)$. The properties of
$\Lambda(x)$, and the consequent constraints on the allowed range
for the deformation parameters, are discussed in
Sec.~\ref{sec:deformedlogarithm}. The deformed algebra arising
from the deformed logarithms and exponential is discussed in
Sec.~\ref{sec:algebra}. Section~\ref{sec:examples} is reserved to
studying specific members of this class: the entropies considered
by  Tsallis, by Abe, and by Kaniadakis. Other up-to-now overlooked
cases are also discussed. The generalized entropies and
distributions related to the deformed logarithms are studied in
Sec.~\ref{sec:entropy}. In Sec.~\ref{sec:lesche} we show that this
two-parametric entropy is stable according to Lesche.
 We summarize the results in the final
section~\ref{sec:conclusion}.

\sect{Canonical formalism} \label{sec:canonical} Guided by the
form of the  BGS entropy Eq.~(\ref{BGSentropy}), we consider the
following class of trace-form entropies
\begin{equation}
S(p)=-\sum_{i=1}^N p_{i}\,\Lambda(p_{i}) \ ,\label{defentropy}
\end{equation}
with $p\equiv\{p_{i}\}_{{i=1,\,\ldots,\,N}}$ a discrete
probability distribution; one may think of $\Lambda(x)$ as a
generalization of the logarithm.

We introduce the
entropic functional
\begin{eqnarray}
\!\!\!\!\!{\cal F}[p]=S(p)\!-\!\beta^\prime\left(\sum_{i=1}^N
p_{_i}\!-\!1\!\right)\!-\!\beta\left(\sum_{i=1}^N
E_{_i}\,p_{_i}-U\!\right) \label{funf}
\end{eqnarray}
with $\beta^\prime$ and $\beta$ Lagrange multipliers. Imposing
that ${\cal F}[p]$ be stationary for variations of $\beta^\prime$
and $\beta$ yields
\begin{eqnarray}
\sum_{i=1}^N\,p_{_i}=1
,\hspace{10mm}\sum_{i=1}^N\,E_{_i}\,p_{_i}=U , \label{con}
\end{eqnarray}
which fix the normalization and mean energy for the canonical
ensemble. In addition, if  ${\cal F}[p]$ in  Eq. (\ref{funf}) is
stationary for variations of the probabilities  $p_{_j}$,
\begin{equation}
\frac{\delta}{\delta p_{_j}} \, \, {\cal F}[p] =0 , \label{df}
\end{equation}
one finds
\begin{eqnarray}
\frac{d}{d\,p_{_j}}\left[p_{_j}\,\Lambda(p_{_j})\right]=
 - \beta\left(E_{_j}-\mu\right) ,
\label{eq}
\end{eqnarray}
where  $\mu=-\beta^\prime/\beta$.

Without loss of generality, we can express the probability
distribution  $p_{_j}$ as
\begin{equation}
p_{_j}=\alpha\,{\cal
E}\left(-{\beta\over\lambda}\,\left(E_{_j}-\mu\right)\right)  ,
\label{distribution}
\end{equation}
where $\alpha$ and $\lambda$ are two arbitrary, real and positive
constants, and ${\cal E}(x)$ a still unspecified invertible
function; we have in mind that  ${\cal E}(x)$ be a generalization
of, and in some limit  reduce to, the exponential function.

Inverting Eq. (\ref{distribution}) and plugging it into
Eq. (\ref{eq}), one finds
\begin{eqnarray}
 \frac{d}{d\,p_{_j}}\left[p_{_j}\,\Lambda(p_{_j})\right]
 =\lambda\, {\cal E}^{-1}\left({p_{_j}\over\alpha}\right)  .
\end{eqnarray}
Up to this point, $\Lambda(x)$ and ${\cal E}(x)$ are two unrelated
functions and our only assumption has been that the entropy has
trace form. Now if we require, by analogy with the relation
between the exponential and the logarithm functions, that
 ${\cal E}(x)$ be the inverse function of
$\Lambda(x)$,  ${\cal E}(\Lambda(x))=\Lambda({\cal E}(x))=x$, we
obtain the following differential-functional equation for
$\Lambda(x)$
\begin{equation}
\frac{d}{d\,p_{_j}}\left[p_{_j}\,\Lambda(p_{_j})\right]
=\lambda\,\Lambda\left(\frac{p_{_j}}{\alpha}\right)
,\label{condint}
\end{equation}
previously introduced in ~\cite{Kaniadakis2}. A simple and
important example in this class of equations is obtained with the
choice $\lambda=1$ and $\alpha=e^{-1}$. In this case it is trivial
to verify that the solution of Eq. (\ref{condint}) that satisfies
the boundary conditions $\Lambda(1)=0$ and
$d\,\Lambda(x)/dx\big|_{x=1}=1$ is $\Lambda(p_{j})=\ln p_{j}$ and
the entropy Eq. (\ref{defentropy}) reduces to the BGS entropy
(\ref{BGSentropy}).

In this paper we will study the deformed logarithms $\Lambda(x)$
that are solutions of Eq.~(\ref{condint}), the corresponding
inverse functions (deformed exponentials), and the  entropies that
can be expressed using these deformed logarithms through
Eq.~(\ref{defentropy}).

\subsection{A counterexample}
Since Eq. (\ref{condint}) imposes a strict condition on the form
of the function $\Lambda(x)$, it is natural to ask what happens if
this condition is relaxed and more general forms of deformed
logarithms are considered. It should be clear from the derivation
of  Eq. (\ref{condint}) that, if such more general logarithms,
which do not satisfy Eq.~(\ref{condint}),
 are used to define the
entropy by means of  Eq. (\ref{defentropy}), the corresponding
distributions cannot be written as Eq. (\ref{distribution}) with
the ``exponential'' ${\cal E}(x)$ the inverse of $\Lambda(x)$.
Alternatively, if one wants that the distribution be of the form
in Eq. (\ref{distribution}) with the ``exponential'' ${\cal E}(x)$
the inverse of $\Lambda(x)$, the entropy cannot be Eq.
(\ref{defentropy}).

For instance, let us consider the following family of
generalized logarithms
\begin{eqnarray}
\ln_{_{(\kappa,\,\xi)}}(x)={\rm sgn}(x-1)\,\Big|\ln_{{\{\kappa\}}}
(x)\Big|^\xi \ ,\label{es1}
\end{eqnarray}
where $\ln_{{\{\kappa\}}}(x)$ is the $\kappa$
logarithm~\cite{Kaniadakis1}, which we discuss in
Sec.~\ref{sec:examples}-C. This family depends on two real
parameters $\kappa\in(-1,\,1)$ and $\xi>0$. We observe that Eq.
(\ref{es1}) is a solution of Eq. (\ref{condint}), for suitable
constants $\alpha$ and $\lambda$, only for the case $\xi=1$, with
$\ln_{{(\kappa,\,1)}}(x)\equiv\ln_{{\{\kappa\}}}(x)$.

The inverse function of Eq. (\ref{es1}) is
\begin{eqnarray}
\exp_{{(\kappa,\,\xi)}}(x)=\exp_{{\{\kappa\}}}\left({\rm sgn}(x)
|x|^{1/\xi}\right) \ .\label{ies1}
\end{eqnarray}
For $\xi=1$ Eq. (\ref{ies1}) reduces to the $\kappa$
exponential~\cite{ Kaniadakis1} $\exp_{_{\{\kappa\}}}(x)$, while
it reduces to the stretched exponential for $\kappa=0$. Moreover,
this family of logarithms and exponentials inherits from the
$\kappa$ logarithm and the $\kappa$ exponential the properties
$\ln_{_{(\kappa,\,\xi)}}(1/x)=-\ln_{_{(\kappa,\,\xi)}}(x) $ and
$\exp_{_{(\kappa,\,\xi)}}(-x)\,\exp_{_{(\kappa,\,\xi)}}(x)=1$.

Introducing the entropy
\begin{equation}
\!S_{_{\kappa,\,\xi}}(p)\!=\!-\lambda\sum_{i=1}^N\int\limits_0\limits^{p_{_i}}
\!\!\ln_{_{(\kappa,\,\xi)}}\!\!\left(\frac{x}{\alpha}\right)\!dx+\lambda\!\int\limits_0\limits^1
\!\!\ln_{_{(\kappa,\,\xi)}}\!\!\left(\frac{x}{\alpha}\right)\!dx \
,\label{funf1}
\end{equation}
the variational principle yields
\begin{eqnarray}
p_{_j}=\alpha\,\exp_{_{(\kappa,\,\xi)}}\!\!\left(-\frac{\beta}{\lambda}\,(E_{_j}-\mu)\right)
\ .\label{dis1}
\end{eqnarray}
Equation (\ref{dis1}) becomes  the
$\kappa$-distribution~\cite{Kaniadakis1} for $\xi=1$ and the
stretched exponential distribution for $\kappa=0$; correspondingly
Eq. (\ref{funf1}) reduces to the $\kappa$ entropy
\begin{eqnarray}
S_{\kappa}(p)=-\sum_{i=1}^Np_{_i}\,\ln_{{\{\kappa\}}}(p_{i})
\end{eqnarray}
in the $\xi\rightarrow1$ limit \cite{Kaniadakis2}, whereas in the
$\kappa\rightarrow0$ limit ($\alpha=\lambda=1$) it
reduces to the stretched exponential entropy \cite{Plastino,Abe5}:
\begin{eqnarray}
S_{\xi}(p)=\sum_{i=1}^N\Gamma(1+\xi,\,-\ln p_{i})-\Gamma(1+\xi) \
,\label{strech}
\end{eqnarray}
where $\Gamma(\mu,\,x)$ is the incomplete gamma function of the
second kind and $\Gamma(\mu)=\Gamma(\mu,\,0)$ is the gamma
function \cite{Ryzhik}. Equation (\ref{strech}) demonstrates that
the entropy (\ref{funf1}) has in general a form different from
Eq.~(\ref{defentropy}): $S_{{\kappa,\,\xi}}\neq -\sum_i
p_i\ln_{{(\kappa,\,\xi)}}(p_i)$.

\subsection{Integrals of the functions $\Lambda(x)$ and ${\cal E}(x)$}
The fact that the generalized logarithm $\Lambda(x)$ is a solution
of the differential equation (\ref{condint}) is sufficient to
calculate its integral
\begin{eqnarray}
\int\limits_{x_1}\limits^{x_2} \Lambda(x)\,dx =\frac{
{x_2}\,\Lambda(\alpha\,{x_2})-{x_1}\,\Lambda(\alpha\,{x_1})}{\lambda}
\ .
\end{eqnarray}
From the definition of the generalized exponential ${\cal E}(x)$
as the inverse of the generalized logarithm $\Lambda(x)$ it is
also simple to calculate the integral of ${\cal E}(x)$ with the
change of variable $x= {\cal E}^{-1} (s) = \Lambda(s) $
\begin{eqnarray}
&&\int\limits_{x_1}\limits^{x_2} {\cal
E}(x)\,dx= x_2\,{\cal E}(x_2)-x_1\,{\cal E}(x_1)  \nonumber \\
&&- \frac{{\cal E}(x_2) \, \Lambda\Big(\alpha\, {\cal
E}(x_2)\Big) - {\cal E}(x_1) \,\Lambda\Big(\alpha\, {\cal E}(x_1)
\Big)}{\lambda} \ . \ \ \ \ \ \ \label{integralcalE}
\end{eqnarray}

\sect{Solutions of the differential-functional equation}
\label{sec:solution} In this Section we study the solutions of Eq.
(\ref{condint}), which we rewrite in the following form:
\begin{eqnarray}
{\displaystyle\frac{d}{d\,x}}\left[x\,\Lambda(x)\right]-\lambda\,\Lambda\left(
\frac{x}{\alpha}\right)=0 \ .\label{condif}
\end{eqnarray}
We shall select the solutions of Eq. (\ref{condif}) that satisfy
appropriate boundary conditions and that keep those properties of
the standard logarithms that we judge important even for a
generalized logarithm.

By performing the change of variable
\begin{eqnarray}
x=\exp\left(\frac{t}{\lambda\,\alpha}\right) \ ,\label{old}
\end{eqnarray}
and introducing the function
\begin{eqnarray}
\Lambda(x)={1\over x}\,f(\lambda\,\alpha\,\ln x) \ ,\label{f}
\end{eqnarray}
the homogeneous differential-functional
equation of the first order shown in
Eq. (\ref{condif}) becomes
\begin{eqnarray}
\frac{d\,f(t)}{d\,t}-f(t-t_{_0})=0 \ ,\label{delay}
\end{eqnarray}
with $ t_{_0}=\lambda\,\alpha\,\ln\alpha$. The most general
solution of Eq.~(\ref{delay}), a differential-difference equation
belonging to the class of delay equations \cite{Bellman}, can be
written in the form
\begin{eqnarray}
f(t)=\sum_{i=1}^n\sum_{j=0}^{m_i-1}a_{{ij}}(s_{_1},\,\ldots,\,s_{_n})\,t^j\,
e^{s_{i}\,t}\ ,\label{sol}
\end{eqnarray}
where $n$ is the number of independent solutions $s_{_i}$ of the
characteristic equation
\begin{eqnarray}
s_{_i}-e^{-t_{_0}\,s_{_i}}=0 \ ,\label{char}
\end{eqnarray}
$m_{i}$ their multiplicity [$s_i$ is solution not only of
Eq.~(\ref{char}), but also of its first $m_i-1$ derivatives] and
$a_{{ij}}$ multiplicative coefficients that depend on the
parameters $s_{_i}$. In terms of the original function and
variable the general solution and the characteristic equation are
\begin{eqnarray}
\Lambda(x)&=&\sum_{i=1}^n\sum_{j=0}^{m_i-1}
a_{{ij}}(s_{1},\,\ldots,\,s_{_n})\,\Big[\lambda\,\alpha\,\ln(x)\Big]^j\,
x^{\lambda\,\alpha\,s_{_i}-1} \nonumber \\
&=& \sum_{i=1}^n\sum_{j=0}^{m_i-1}
a^\prime_{{ij}}(\kappa_{1},\,\ldots,\,\kappa_{n})\,\left[\ln(x)\right]^j\,
x^{\kappa_{i}} \ ,\label{solorig}
\end{eqnarray}
\begin{eqnarray}
1+\kappa_{_i}=\lambda\,\alpha^{-\kappa_{_i}} \ ,\label{condition}
\end{eqnarray}
where $\kappa_{i}=\lambda\,\alpha\,s_{i}-1$ and
$a^\prime_{{ij}}=(\lambda\,\alpha)^j\,a_{{ij}}$.

In the present work, we are interested in nonoscillatory solutions
for Eq.~(\ref{delay}): this kind of solutions maintain a closer
relation with the standard logarithm. Therefore, we consider only
real solutions of Eq.~(\ref{char}). There exist four different
cases depending on the value of  $ t_{_0}$: (a) for $t_{_0}\geq0$
we have one solution, $n=1$ and $m=1$; (b) for $-1/e<t_{_0}<0$, we
have two nondegenerate solutions, $n=2$ and $m_{_i}=1$; (c) for
$t_{_0}=-1/e$ we have two degenerate solutions, $n=1$ and $m=2$;
(d) for $t_{_0}<-1/e$ there exist no solutions.

We discuss in order the three cases (a), (b), and (c) that
yield solutions of the delay equation (\ref{delay}) and, therefore,
of the corresponding Eq. (\ref{condif}).

The case (a) is the least interesting:
\begin{eqnarray}
\Lambda(x)=a\,x^\kappa
\end{eqnarray}
is just a single power and cannot change sign as one would require
from a logarithm.

In the case (b), we  obtain a binomial solution:
\begin{eqnarray}
\Lambda(x)=A_{_1}(\kappa_{_1},\,\kappa_{_2})\,
x^{\kappa_{_1}}+A_{_2}(\kappa_{_1},\,\kappa_{_2})\,
x^{\kappa_{_2}}\ ,\label{sol1}
\end{eqnarray}
where $A_{_1}$ and $A_{_2}$ are the integration constants.

The characteristic equation~(\ref{condition}) can be solved for
the two constants $\alpha$ and $\lambda$,
\begin{eqnarray}
\alpha =
\left(\frac{1+\kappa_{_2}}{1+\kappa_{_1}}\right)^{1/(\kappa_{_1}
-\kappa_{_2})} \ ,
\end{eqnarray}
\begin{eqnarray}
\lambda&=&\frac{(1+\kappa_{_2})}{\mbox{\raisebox{-1mm}
{$(1+\kappa_{_1})$}}}^{\scriptscriptstyle{\kappa_{_1}/(\kappa_{_1}-\kappa_{_2})}}
_{\mbox{\raisebox{3.5mm}{$\scriptscriptstyle{\kappa_{_2}/(\kappa_{_1}-\kappa_{_2})}$}}}
 \ .
\end{eqnarray}
The two arbitrary coefficients $A_{_1}$ and $A_{_2}$ correspond to
the freedom of scaling $x$ and $\Lambda(x)$ in Eq.~(\ref{condif}).
We fix these integration constants using the two boundary
conditions
\begin{eqnarray}
&&\Lambda(1)=0 \ , \label{norm}
\\ &&\frac{d\,\Lambda(x)}{d\,x}\Bigg|_{x=1}\!\!\!=1 \ .\label{slope}
\end{eqnarray}
The first condition implies $A_1=-A_2\equiv A$ while from the
second one has $A=1/(\kappa_{_1}-\kappa_{_2})$.  Equation
(\ref{sol1}) assumes the final expression
\begin{eqnarray}
\Lambda(x)=\frac{x^{\kappa_{_1}}-x^{\kappa_{_2}}}{\kappa_{_1}
-\kappa_{_2}} \ ,\label{log1}
\end{eqnarray}
a two-parameter function which reduces to the standard logarithm
in the $(\kappa_{_1},\,\kappa_{_2})\rightarrow(0,\,0)$ limit.

After introducing the notation
$\Lambda(x)=\ln_{_{\{\kappa,\,r\}}}(x)$ and the two auxiliary
parameters  $\kappa=(\kappa_{_1}-\kappa_{_2})/2$ and
$r=(\kappa_{_1}+\kappa_{_2})/2$, Eq.~(\ref{log1}) becomes
\begin{eqnarray}
\nonumber \ln_{_{\{\kappa,\,r\}}}(x)&=&
x^r\,{\displaystyle\frac{x^\kappa-x^{-\kappa}}{2\,\kappa}}\\
&=&x^r\,\ln_{_{\{\kappa\}}}(x) \ . \label{log}
\end{eqnarray}

The constants $\alpha$ and $\lambda$ expressed in terms of
$\kappa$ and $r$
\begin{eqnarray}
\alpha=\left(\frac{1+r-\kappa}{1+r+\kappa}\right)^{1/(2\kappa)} \
,\label{al}
\end{eqnarray}
\begin{eqnarray}
\lambda=\frac{\left(1+r-\kappa\right)}{\mbox{\raisebox{-1mm}
{$\left(1+r+\kappa\right)$}}}
^{\scriptscriptstyle{(r+\kappa)/(2\kappa)}}_
{\mbox{\raisebox{3.5mm}{$\scriptscriptstyle{(r-\kappa)/(2\kappa)}$}}}
 \ ,\label{l}
\end{eqnarray}
are symmetric for $\kappa\leftrightarrow-\kappa$ and satisfy the
useful relations $(1+r \pm \kappa)\,\alpha^{r \pm
\kappa}=\lambda$, which is the characteristic
equation~(\ref{condition}), and
$1/\lambda=\ln_{_{\{\kappa,\,r\}}}(1/\alpha)$, which is the
differential equation~(\ref{condif}) at $x=1$. In the following,
we call the solution (\ref{log}) the deformed logarithm or
$(\kappa,\,r)$ logarithm.

Finally, we consider the case (c). This case can be also
obtained as the limit of case (b) when
the two distinct solutions $s_{_1}$ and $s_{_2}$ become
degenerate. When $t_{_0}=-1/e$,
$s=e$ verifies not only
Eq. (\ref{char}), but also its first derivative
$1+t_0 \exp(-t_0 s)=0$: this solution is twice degenerate
($s_1=s_2=e$); Eq. (\ref{solorig})
with the boundary
conditions (\ref{norm}) and (\ref{slope}) becomes
\begin{eqnarray}
\Lambda(x)=x^{r}\,\ln x \ ,\label{part}
\end{eqnarray}
where the parameter $r=\lambda\,\alpha\,e-1$. The standard
logarithm $\Lambda(x)=\ln x$ is recovered for $r=0$; the same
standard logarithm is actually recovered from the $(\kappa,\,r)$
logarithm in the limit $(\kappa,\,r) \to (0,\,0)$ independently of
the direction.

\sect{Properties of deformed functions}
\label{sec:deformedlogarithm}
The properties of the entropy (\ref{defentropy}), and of the
corresponding distribution (\ref{distribution}), follow from the
properties of the deformed logarithm
$\Lambda(x)\equiv\ln_{_{\{{\scriptstyle \kappa,\,r}\}}}(x)$, which
is used in its definition. Naudts~\cite{Naudts1} gives a list of
general properties that a deformed logarithm must satisfy in order
that the ensuing entropy and distribution function be physical. In
this section we determine the region of parameter space $(\kappa,
r)$ where the logarithm (\ref{log}) satisfies these properties and
list the corresponding properties of its inverse, the $(\kappa,
r)$ exponential.

\subsection{$\bfm{(\kappa,\,r)}$-deformed logarithm}

\begin{figure}[ht]
\includegraphics[angle=0,width=\columnwidth]{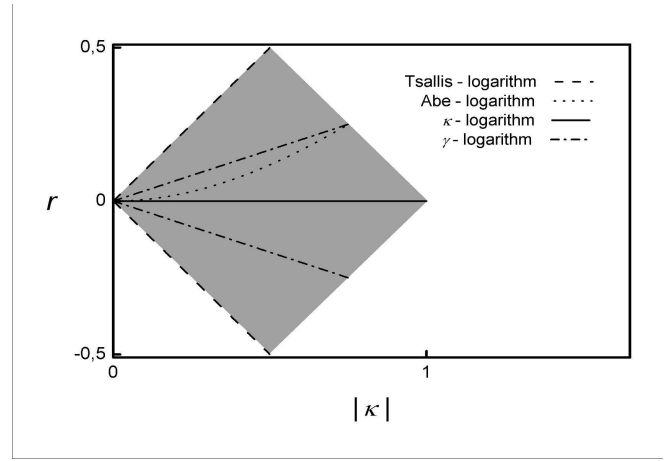}
\caption{Parameter space $(\kappa, r)$ for the logarithm
(\ref{log}). The shaded region represents the constraints of Eq.
(\ref{quadri}) on the parameters. The four lines, dashed, dotted,
solid, and dash-dotted, correspond to the Tsallis (\ref{tlog}),
Abe (\ref{abelog}), $\kappa$ (\ref{klog}), and $\gamma$
(\ref{alog}) logarithm, respectively. \label{fig:parameterRegion}
}
\end{figure}

The following properties for the $(\kappa,\,r)$ logarithm hold
when $\kappa$ and $r$ satisfy
 the corresponding limitations:
\begin{eqnarray}
&&\ln_{_{\{{\scriptstyle \kappa,\,r}\}}}(x) \in C^{\infty}(I\!\!
R^+) \ ,
\label{pd1}\\
&&\frac{d}{d\,x}\, \ln_{_{\{{\scriptstyle \kappa,\,r}\}}}(x)>0 \
,\hspace{2mm}-|\kappa|\leq r\leq|\kappa| \ ,
\label{pd2}\\
&&\frac{d^2}{d\,x^2}\, \ln_{_{\{{\scriptstyle
\kappa,\,r}\}}}(x)<0 \ ,\hspace{1mm}\, -|\kappa|\leq
r\leq\frac{1}{2}-\Big|\frac{1}{2}-|\kappa|\Big| \ ,
\label{pd3}\\
&&\ln_{_{\{{\scriptstyle \kappa,\,r}\}}}(1)=0 \ , \label{pd4}\\
&&\int\limits_0\limits^1\ln_{_{\{\kappa,\,r\}}}(x)\,dx=-\frac{1}{(1+r)^2
-\kappa^2} \ ,\hspace{1mm}1+r>|\kappa| \ ,\label{pd5} \\
&&\int\limits_0\limits^1\ln_{_{\{\kappa,\,r\}}}
\!\!\left(\frac{1}{x}\right)\,dx=\frac{1}{(1-r)^2 -\kappa^2} \
,\hspace{1mm}1-r>|\kappa| \ . \ \ \ \ \ \ \label{pd51}
\end{eqnarray}
Equation (\ref{pd1}) states that the $(\kappa,\,r)$ logarithm is
an analytical function for all $x\geq0$ and for all $\kappa,\,r\in
I\!\!R$; Eq. (\ref{pd2}) that it is a strictly increasing function
for $-|\kappa|\leq r\leq|\kappa|$; Eq. (\ref{pd3}) that it is
concave for $-|\kappa|\leq r\leq|\kappa|$, when $|\kappa|<1/2$,
and for $-|\kappa|\leq r<1-|\kappa|$, when $|\kappa|\geq 1/2$; Eq.
(\ref{pd4}) states that the  $(\kappa,\,r)$ logarithm satisfies
the boundary condition (\ref{norm}); Eqs. (\ref{pd5}) and
(\ref{pd51}) that it has at most integrable divergences for
$x\to0^+$ and $x\to +\infty$. These two last conditions
(\ref{pd5}) and (\ref{pd51}) assure the normalization of the
canonical ensembles distribution arising from entropy
(\ref{defentropy}). Conditions (\ref{pd2})-(\ref{pd51}) select the
following region:
\begin{eqnarray}
I\!\!R^2\supset{\cal R}=\left\{
\begin{array}{l}
\hspace{4.5mm}-|\kappa|\leq r\leq|\kappa|\hspace{5mm} {\rm if} \
0\leq|\kappa|<{1\over2} \\
|\kappa|-1<r<1-|\kappa|\hspace{5mm} {\rm if} \
{1\over2}\leq|\kappa|<1
\end{array} \ \
\right.\label{quadri}
\end{eqnarray}
which is shown in Fig.~\ref{fig:parameterRegion}: every point in
$\cal R$ selects a deformed logarithm that satisfies all
properties (\ref{pd1})-(\ref{pd51}). This  region $\cal R$ include
values of the parameters for which the  logarithm  is finite in
the limit $x\rightarrow0$ or $x\rightarrow+\infty$. Note that
points in region (\ref{quadri}) always satisfy $|\kappa|<1$, the
condition obtained in Ref. \cite{Kaniadakis2} for the case $r=0$.
In addition, $\kappa\rightarrow0$ implies $r\rightarrow0$.
Following Ref. \cite{Naudts1} we introduce the dual logarithm
\begin{eqnarray}
\ln_{_{\{\kappa,\,r\}}}^\ast(x)=-\ln_{_{\{\kappa,\,r\}}}\!\!\left(\frac{1}{x}
\right) \ ,\label{dual}
\end{eqnarray}
which is related to the original $(\kappa,\,r)$ logarithm by
\begin{equation}
\ln_{{\{\kappa,\,r\}}}^\ast(x)=\ln_{{\{\kappa,\,-r\}}}(x)
=x^{-2\,r}\ln_{{\{\kappa,r\}}}(x) .\label{pd6}
\end{equation}
Note  that this last result implies that Eq. (\ref{pd51}) is
equivalent to Eq. (\ref{pd5}) on exchanging $r\leftrightarrow-r$.
Let us define that a logarithm is  self-dual when
\begin{eqnarray}
\ln_{_{\{\kappa,\,r\}}}(x)=\ln_{_{\{\kappa,\,r\}}}^\ast(x) \
.\label{self}
\end{eqnarray}
Then  Eq. (\ref{pd6}) shows that $\ln_{_{\{\kappa,\,r\}}}(x)$ is
self-dual if and only if $r=0$: this case coincides with the
$\kappa$ logarithm \cite{Kaniadakis1,Kaniadakis2}.

The asymptotic behavior of $\ln_{_{\{\kappa,r\}}}(x)$ for $x$
approaching zero is
\begin{eqnarray}
\ln_{_{\{\kappa,\,r\}}}(x){\atop\stackrel{\textstyle\sim}{\scriptstyle
x\rightarrow
{0^+}}}-\frac{1}{2\,|\kappa|}\,\frac{1}{x^{|\kappa|-r}} \
;\label{pd7}
\end{eqnarray}
in particular it results that $\ln_{_{\{{\scriptstyle
\kappa,\,r}\}}}(0^+)=-\infty$  for $r<|\kappa|$,  while, if
$r=|\kappa|$, $\ln_{_{\{\kappa,\,|\kappa|\}}}(0^+)=-1/2\,|\kappa|$
is finite. Consequently, its inverse
$\exp_{_{\{\kappa,\,|\kappa|\}}}(x)$ goes to zero at finite $x$:
the distribution function has a cut-off at finite energy,
$\exp_{_{\{\kappa,\,|\kappa|\}}}(-1/2\,|\kappa|)=0$.\\
Analogously, the behavior of $\ln_{_{\{\kappa,\,r\}}}(x)$ for
large values of $x$ is
\begin{eqnarray}
\ln_{_{\{\kappa,\,r\}}}(x){\atop\stackrel{\textstyle\sim}{\scriptstyle
x\rightarrow {+\infty}}}\frac{x^{|\kappa|+r}}{2\,|\kappa|} \
,\label{pd8}
\end{eqnarray}
which implies $\ln_{_{\{{\scriptstyle
\kappa,\,r}\}}}(+\infty)=+\infty$ for $r>-|\kappa|$, while again,
if $r=-|\kappa|$, then $ \ln_{_{\{{\scriptstyle
\kappa,\,-|\kappa|}\}}}(+\infty) = 1/2\,|\kappa|$ and
 $\exp_{_{\{\kappa,-|\kappa|\}}}(1/2\,|\kappa|)=+\infty$.

The generalized logarithm verifies the following scaling law
\begin{eqnarray}
a\,\ln_{_{\{\kappa,\,r\}}}(x)=\ln_{_{\{\kappa/a,\,r/a\}}}(x^a) \
,\label{resclog}
\end{eqnarray}
of which  Eq. (\ref{pd6}) is a particular case for $a=-1$, and
that becomes   $a\,\ln x=\ln(x^a)$ when
$(\kappa,\,r)\rightarrow(0,\,0)$.

In the following we gives some useful relations. First the
relation
\begin{eqnarray}
\ln_{_{\{\kappa,\,r\}}}(x\,y)&=&{1\over2}\left(x^{r+\kappa}+x^{r-\kappa}\right)
\,\ln_{_{\{\kappa,\,r\}}} (y)\nonumber \\
&+&{1\over2}\left(y^{r+\kappa}+y^{r-\kappa}\right)\,\ln_{_{\{\kappa,\,r\}}}
(x)   \label{rl1}
\end{eqnarray}
is easily proved taking into account the definition (\ref{log}).
By using the identity
$y^{r-\kappa}=y^{r+\kappa}-2\,\kappa\,\ln_{_{\{\kappa,\,r\}}}(y)$,
Eq. (\ref{rl1}) becames
\begin{eqnarray}
\ln_{_{\{\kappa,\,r\}}}(x\,y)&=&x^{r+\kappa}\,\ln_{_{\{\kappa,\,r\}}}
(y)+y^{r+\kappa}\,\ln_{_{\{\kappa,\,r\}}}(x) \nonumber \\
&-&2\,\kappa\,\ln_{_{\{\kappa,\,r\}}}(x)\,\ln_{_{\{\kappa,\,r\}}}(y)
\ . \ \ \ \ \ \ \ \ \ \ \label{logpro}
\end{eqnarray}

Moreover, using
\begin{equation}
{1\over2}\left(x^{r+\kappa}\!+\!x^{r-\kappa}\right)=-(1+r)\,\ln_{_{\{\kappa,\,r\}}}
\!(x)\nonumber +\lambda\,\ln_{_{\{\kappa,\,r\}}}
\!\left({x/\alpha}\right) \ ,
\end{equation}
Eq. (\ref{rl1}) can be rewritten as
\begin{eqnarray}
\nonumber
\ln_{_{\{\kappa,\,r\}}}(x\,y)&=&-2\,(1+r)\,\ln_{_{\{\kappa,\,r\}}}
(x)\,\ln_{_{\{\kappa,\,r\}}}
(y)\\
&+&\lambda\, \ln_{_{\{\kappa,\,r\}}} (x)\,\ln_{_{\{\kappa,\,r\}}}
\left({y\over\alpha}\right)\nonumber \\ &+& \lambda\,
\ln_{_{\{\kappa,\,r\}}}
\left({x\over\alpha}\right)\,\ln_{_{\{\kappa,\,r\}}} (y) \
.\label{sl}
\end{eqnarray}

\subsection{$\bfm{(\kappa,\,r)}$-deformed exponential}

The deformed logarithm is a strictly increasing function for
$-|\kappa|\leq r \leq |\kappa|$; therefore, it can be inverted for
$(\kappa,\,r)\in{\cal R}$. We call its inverse the deformed
exponential $\exp_{_{\{{\scriptstyle \kappa,\,r}\}}}(x)$, whose
analytical properties follow from those of the deformed logarithm:
\begin{eqnarray}
&&\exp_{_{\{{\scriptstyle \kappa,\,r}\}}}(x) \in
C^{\infty}(I\!\!I) \quad ,
\label{ed1} \\
&&\frac{d}{d\,x}\, \exp_{_{\{{\scriptstyle \kappa,\,r}\}}}(x)>0 \
, \label{ed2}\\
&&\frac{d^2}{d\,x^2}\, \exp_{_{\{{\scriptstyle
\kappa,\,r}\}}}(x)>0 \ ,
\label{ed3}\\
&&\exp_{_{\{{\scriptstyle
\kappa,\,r}\}}}(0)=1 \ ,
\label{ed4}\\
&&\int\limits_{-\infty}\limits^0\exp_{_{\{\kappa,\,r\}}}(x)\,dx=\frac{1}
{(1+r)^2-\kappa^2} \ , \label{ed5} \\
&&\int\limits_{-\infty}\limits^0{dx\over\exp_{_{\{\kappa,\,r\}}}(-x)}=\frac{1}
{(1-r)^2-\kappa^2} \ .\label{ed51}
\end{eqnarray}
Equation (\ref{ed1}) states that  the deformed exponential
$\exp_{_{\{\kappa,\,r\}}}(x)$ is a continuous function for all
$x\in I\!\!I$, where $I\!\!I=I\!\!R^+$, when $-|\kappa|<r<
|\kappa|$, $I\!\!I=(-1/2\,|\kappa|,\,\infty)$, when $r =
|\kappa|$, and $I\!\!I=(-\infty,\,1/2\,|\kappa|)$, when $r =
-|\kappa|$. Equations (\ref{ed2})-(\ref{ed51}) state that
$\exp_{_{\{\kappa,\,r\}}}(x)$ is a strictly increasing and convex
function, normalized according to Eq.~(\ref{ed4}), and which goes
to zero fast enough to be integrable for $x \to\pm\infty$.

Introducing the dual of the exponential function
\begin{eqnarray}
\exp_{_{\{\kappa,\,r\}}}^\ast(x)={1\over\exp_{_{\{\kappa,\,r\}}}(-x)}
\ ,
\end{eqnarray}
Eq. (\ref{pd6}) implies
\begin{eqnarray}
\exp_{_{\{\kappa,\,r\}}}^\ast(x)=\exp_{_{\{\kappa,\,-r\}}}(x) \ ,
\end{eqnarray}
which means
\begin{eqnarray}
\exp_{_{\{\kappa,\,r\}}}(x)\,\exp_{_{\{\kappa,\,-r\}}}(-x)=1 \
.\label{ed76}
\end{eqnarray}
Only when $r=0$ does this relation reproduce that of the standard
exponential \cite{Kaniadakis2}.

The asymptotic behaviors  (\ref{pd7}) and (\ref{pd8}) of
$\ln_{_{\{\kappa,\,r\}}}(x)$ imply
\begin{eqnarray}
\exp_{_{\{\kappa,\,r\}}}(x){\atop\stackrel{\textstyle\sim}{\scriptstyle
x\rightarrow {\pm\infty}}}|2\,\kappa\,x|^{1/(r\pm|\kappa|)} \
;\label{expasy}
\end{eqnarray}
in particular
\begin{eqnarray}
&&\exp_{_{\{{\scriptstyle \kappa,\,r}\}}}(-\infty)=0^+ \
,\hspace{5mm}{\rm for}\hspace{5mm}r<|\kappa| \ ,
\label{ed7}\\
&&\exp_{_{\{{\scriptstyle \kappa,\,r}\}}}(+\infty)=+\infty \
,\hspace{5mm}{\rm for}\hspace{5mm}r>-|\kappa| \ , \ \ \ \ \
\label{ed8}
\end{eqnarray}
while
\begin{eqnarray}
&&\exp_{_{\{{\scriptstyle \kappa,\,r}\}}}(-1/2\,|\kappa|)=0^+ \
,\hspace{3mm}{\rm when}\hspace{3mm}r=|\kappa| \ ,
\label{ed7b}\\
&&\exp_{_{\{{\scriptstyle \kappa,\,r}\}}}(+1/2\,|\kappa|)=+\infty
\ ,\hspace{3mm}{\rm when}\hspace{3mm}r=-|\kappa| \ . \ \ \ \ \ \
\ \  \label{ed8b}
\end{eqnarray}

Finally, the scaling law
\begin{eqnarray}
\left[\exp_{_{\{\kappa,\,r\}}}(x)\right]^a=\exp_{_{\{\kappa/a,\,r/a\}}}
(a\,x)\ ,\label{rescexp}
\end{eqnarray}
reduces to Eq. (\ref{ed76}) for $a=-1$ and reproduces the property
$[\exp(x)]^a=\exp(a\,x)$ in the $(\kappa,\,r)\rightarrow(0,\,0)$
limit.


\sect{Deformed algebra} \label{sec:algebra} Using the definition
of the deformed logarithm and its inverse function, we can
introduce two composition laws, the deformed sum
$x\oplus\mbox{\raisebox{3mm}{\hspace{-4.5mm}$\scriptscriptstyle
{\kappa,r}$}} \hspace{1mm}y$ and product
$x\otimes\mbox{\raisebox{-2mm}{\hspace{-4.5mm}$\scriptscriptstyle
{\kappa,r}$}} \hspace{1mm}y$.

Let us define the deformed sum:
\begin{eqnarray}
x\oplus\mbox{\raisebox{3mm}{\hspace{-4.5mm}$\scriptscriptstyle
{\kappa,r}$}}
\hspace{1mm}y=\ln_{_{\{\kappa,\,r\}}}\left(\exp_{_{\{\kappa,r\}}}(x)\,
\exp_{_{\{\kappa,r\}}}(y)\right) \ ,\label{sd1}
\end{eqnarray}
which reduces, in the $(\kappa,\,r)\rightarrow(0,\,0)$ limit, to
the ordinary sum
$x\oplus\mbox{\raisebox{3mm}{\hspace{-4mm}$\scriptscriptstyle
{0,0}$}} \hspace{1mm}y=x+y$. Its definition implies that the
deformed sum satisfies the following properties:\\
(a) is associative; (b) is commutative; (c) its neutral element
is $0$;  (d) the opposite of $x$ is $ \ln_{_{\{\kappa,\,r\}}}
\!\!\left(1/\exp_{_{\{\kappa,r\}}}(x)\right)$.

If $x$ and $y$ are positive Eq. (\ref{sd1}) yields
\begin{eqnarray}
\ln_{_{\{\kappa,\,r\}}}(x\,y)=\ln_{_{\{\kappa,\,r\}}}
(x)\oplus\mbox{\raisebox{3mm}{\hspace{-4.5mm}$\scriptscriptstyle
{\kappa,r}$}} \hspace{1mm}\ln_{_{\{\kappa,\,r\}}}(y) \
,\label{sumlog}
\end{eqnarray}
which, when $(\kappa,\,r)\rightarrow(0,\,0)$, reduces to the
well-known property $\log\,(x\,y)=\log x+\log y$.

In the same way, let us introduce the deformed product between
positive $x$ and $y$:
\begin{eqnarray}
x\otimes\mbox{\raisebox{-2mm}{\hspace{-4.5mm}$\scriptscriptstyle
{\kappa,r}$}}
\hspace{1mm}y=\exp_{_{\{\kappa,r\}}}\left(\ln_{_{\{\kappa,\,r\}}}(x)+
\ln_{_{\{\kappa,\,r\}}}(y)\right) \ ,\label{prd1}
\end{eqnarray}
which reduces, for $(\kappa,\,r)\rightarrow(0,\,0)$, to the
ordinary product
$x\otimes\mbox{\raisebox{-2mm}{\hspace{-4.5mm}$\scriptscriptstyle
{0,0}$}} \hspace{1mm}y=x\,y$. This product verifies the following
properties: (a) is associative; (b) is commutative; (c) its
neutral element is $1$; (d) the inverse element of $x$ is
$\exp_{_{\{\kappa,r\}}} \left(-\ln_{_{\{\kappa,\,r\}}}(x)\right)$.

According to Eq. (\ref{prd1}) we have
\begin{eqnarray}
\exp_{_{\{\kappa,r\}}}(x+y)=
\exp_{_{\{\kappa,r\}}}(x)\otimes\mbox{\raisebox{-2mm}{\hspace{-4.5mm}$\scriptscriptstyle
{\kappa,r}$}}\hspace{1mm}\exp_{_{\{\kappa,r\}}}(y) \
,\label{sumexp}
\end{eqnarray}
which reproduces in the $(\kappa,\,r)\rightarrow(0,\,0)$ limit the
well-know property of the exponential $\exp (x)\,\exp
(y)=\exp(x+y)$.

Note that the algebraic structures
$A_{_1}\equiv(I\!\!R,\,\oplus\mbox{\raisebox{3mm}{\hspace{-3.4mm}$\scriptscriptstyle
{\kappa,r}$}} \hspace{1mm})$ and
$A_{_2}\equiv(I\!\!R^+,\,\otimes\mbox{\raisebox{-2mm}{\hspace{-3.5mm}$\scriptscriptstyle
{\kappa,r}$}} \hspace{1mm})$ are two Abelian groups. The deformed
sum (\ref{sd1}) and product (\ref{prd1}) are not distributive and
the structure
$A_{_3}\equiv(I\!\!R^+,\,\oplus\mbox{\raisebox{3mm}{\hspace{-3.4mm}$\scriptscriptstyle
{\kappa,r}$}}
\hspace{1mm},\,\otimes\mbox{\raisebox{-2mm}{\hspace{-3.5mm}$\scriptscriptstyle
{\kappa,r}$}} \hspace{1mm})$ is not an Abelian field. In any case,
following Ref. \cite{Kaniadakis2} it is possible to define a
deformed product
$\otimes\mbox{\raisebox{3mm}{\hspace{-3.4mm}$\scriptscriptstyle
{\kappa,r}$}}$ and sum
$\oplus\mbox{\raisebox{-2mm}{\hspace{-3.5mm}$\scriptscriptstyle
{\kappa,r}$}}$ which are distributive with respect to
$\oplus\mbox{\raisebox{3mm}{\hspace{-3.4mm}$\scriptscriptstyle
{\kappa,r}$}}$ and
$\otimes\mbox{\raisebox{-2mm}{\hspace{-3.5mm}$\scriptscriptstyle
{\kappa,r}$}}$, respectively, so that the structures ${\cal
A}_1\equiv(I\!\!R^+,\,\oplus\mbox{\raisebox{3mm}{\hspace{-3.4mm}$\scriptscriptstyle
{\kappa,r}$}},\,\otimes\mbox{\raisebox{3mm}{\hspace{-3.4mm}$\scriptscriptstyle
{\kappa,r}$}} \hspace{1mm})$ and ${\cal
A}_2\equiv(I\!\!R^+,\,\oplus\mbox{\raisebox{-2mm}{\hspace{-3.5mm}$\scriptscriptstyle
{\kappa,r}$}},\,\otimes\mbox{\raisebox{-2mm}{\hspace{-3.5mm}$\scriptscriptstyle
{\kappa,r}$}} \hspace{1mm})$ are Abelian.

Finally, from Eqs. (\ref{sumlog}) and (\ref{sl}), we obtain
\begin{equation}
x\oplus\mbox{\raisebox{3mm}{\hspace{-4mm}$\scriptscriptstyle
{\kappa,r}$}}
\hspace{1mm}y=x\left[\exp_{_{\{\kappa,\,r\}}}(y)\right]^{r+\kappa}
+y\left[\exp_{_{\{\kappa,\,r\}}}(x)\right]^{r+\kappa}
-2\,\kappa\,x\,y  .\label{sd2}
\end{equation}
From the practical point of view this last expression, like all
the expressions involving the $(\kappa,\,r)$ exponentials, are
more useful for those particular values of $r$ and $\kappa$ for
which an explicit closed form of the $(\kappa,\,r)$ exponential
can be given. In the next section we shall see some examples.


\sect{Examples of one-parameter deformed logarithms}
\label{sec:examples}
The  two-parameter class of deformed logarithms~(\ref{log})
includes an infinity of one-parameter deformed logarithms that can
be specified by selecting a relation between $\kappa$ and $r$. In
this section we discuss a few specific one-parameter logarithms
that are already known in the literature and have been used to
define entropies  in the context of generalizations of statistical
mechanics and thermodynamics: we show that they are in fact
members of the same two-parameter class; we also  introduce a few
different examples of one-parameter logarithms.

\subsection{Tsallis logarithm}\label{subsec:Tsallis}
The first example is obtained with the choice $r=-\kappa$ for $
-1/2<\kappa<1/2$. After introducing the parameter $q=1+2\kappa$
[$0< q <2$] we obtain the Tsallis logarithm $\ln_{q}(x) \equiv
\ln_{{\{(q-1)/2\,,\,\,(1-q)/2\}}}(x)$ and the Tsallis exponential
 $\exp_{q}(x)$ as follows:
\begin{eqnarray}
&&\ln_{q}(x)=\frac{x^{1-q}-1}{1-q} \ , \label{tlog} \\
&&\exp_{q}(x)= [1+(1-q)\,x]^{1/(1-q)} \ . \label{qexp}
\end{eqnarray}
The relation (\ref{pd6}) reads
\begin{eqnarray}
\ln_{_{q}}\,(x)=-\ln_{_{2-q}}\,\left(\frac{1}{x}\right) \
,\label{condition1}
\end{eqnarray}
while Eq. (\ref{logpro}) becomes
\begin{eqnarray}
&&\!\!\!\ln_{_q}(x\,y)=\ln_{_q}(x)\!+\!\ln_{_q}(y)\!+\!(1\!-\!q)\,
\ln_{_q}(x)\,\ln_{_q}(y) \ . \ \ \ \ \
\end{eqnarray}
The $q$-deformed algebra already discussed in
Refs.~\cite{Borges2,Wang} results as a particular case with
$r=-\kappa=(1-q)/2$ of the deformed algebra discussed in
Sec.~\ref{sec:algebra}.

\subsection{Abe logarithm}\label{subsec:Abe}
As a second example we consider  the constraint
$(r+1)^2=1+\kappa^2$ and define $q_{_{\rm A}}=r+\kappa+1$. Then
the two-parameter logarithm in Eq.~(\ref{log}) becomes the
logarithm associated with the entropy introduced by Abe
\cite{Abe3}
\begin{eqnarray}
\ln_{q_{_{\rm A}}}\,(x)=\frac{x^{(q_{_{\rm
A}}^{-1})-1}-x^{q_{_{\rm A}}-1}}{q_{_{\rm A}}^{-1}-q_{_{\rm A}}} \
,\label{abelog}
\end{eqnarray}
which reduces to the standard logarithm for $q_{_{\rm
A}}\rightarrow1$. The invariance of Eq.~(\ref{log}) for $\kappa\to
-\kappa$ results in $\ln_{q_{_{\rm A}}}(x)$ being invariant for
$q_{_{\rm A}}\to 1/q_{_{\rm A}}$. In this case the inverse
function of the Abe logarithm~(\ref{abelog}), which exists because
$\ln_{q_{{\rm A}}}\,(x)$ is monotonic for $1/2<q_{_{\rm A}}<2$,
cannot be express in terms of elementary functions, since
Eq.~(\ref{abelog}) is not invertible algebraically.  We remark
that  Eq.~(\ref{logpro}) in the present case reads
\begin{eqnarray}
\ln_{q_{_{\rm A}}}(x\,y)&=&x^{q_{_{\rm A}}-1}\,\ln_{q_{_{\rm
A}}}(y)+y^{q_{_{\rm A}}-1}\,\ln_{q_{_{\rm A}}}(x) \nonumber \\
&+&\left(q_{_{\rm A}}^{-1}-q_{_{\rm A}}\right)\, \ln_{q_{_{\rm
A}}}(x)\,\ln_{q_{_{\rm A}}}(y) \ .
\end{eqnarray}

\subsection{$\kappa$ logarithm}\label{subsec:Kaniadakis}
Our third example is obtained with the constraint $r=0$.
Introducing the notation $\ln_{_{\{\kappa\}}}(x) \equiv
\ln_{_{\{\kappa,0\}}}(x) $ from Eq. (\ref{log}) we obtain the
$\kappa$ logarithm and consequently its inverse function, namely,
the $\kappa$ exponential introduced in
\cite{Kaniadakis1,Kaniadakis2}:
\begin{eqnarray}
&&\ln_{_{\{\kappa\}}}(x)=\frac{x^\kappa-x^{-\kappa}}{2\,\kappa} \
,\label{klog} \\
&&\exp_{_{\{\kappa\}}}(x)=\left(\sqrt{1+\kappa^2\,x^2}+\kappa\,x
\right)^{1/\kappa}\ ,\label{kexp}
\end{eqnarray}
with $\kappa\in(-1,\,1)$. We remind the reader that, because of
property~(\ref{pd6}), the $\kappa$ logarithm is the only member of
the family that is self-dual
\begin{eqnarray}
\ln_{_{\{\kappa\}}}(x)=-\ln_{_{\{\kappa\}}}\left(\frac{1}{x}\right)
\ .
\end{eqnarray}

The function $\exp_{_{\{\kappa\}}}(x)$ increases at the same rate
that  the function $\exp_{_{\{\kappa\}}}(-x)$ decreases,
\begin{eqnarray}
\exp_{_{\{\kappa\}}}(x)\,\exp_{_{\{\kappa\}}}(-x)=1 \ .
\end{eqnarray}

The $\kappa$-deformed sum is obtained from the more general
Eq.~(\ref{sd1}) by setting $r=0$:
\begin{eqnarray}
x\oplus\!\!\!\!\!^{^{\scriptscriptstyle \kappa}}\,\,
y=x\,\sqrt{1+\kappa^2\,y^2}+y\,\sqrt{1+\kappa^2\,x^2_{ }} \
,\label{ksum}
\end{eqnarray}
which reduces to the ordinary sum for $\kappa\rightarrow0$. The
opposite approach, \emph{i.e.}, starting from the
$\kappa$-deformed sum (\ref{ksum}) to obtain the $\kappa$
logarithm and $\kappa$ exponential has been taken in
Refs.~\cite{Kaniadakis1,Kaniadakis2}, where it is shown that the
$\kappa$-deformed sum is the additivity law of the relativistic
momenta.

\subsection{Other examples}\label{subsec:gamma}
If we define the parameter $w=r/|\kappa|$,  we observe that when
$w=0,\,\pm1/3,\,\pm1/2,\,\pm1,\,\pm5/3,\,\pm2,\,\pm3,\,\pm5$, and
$\pm7$, the inverse function of the deformed logarithm can be
found by solving an algebraic equation of degree not larger than
4; the corresponding deformed exponential can be written
explicitly. In particular, the cases $w=0$ and $\pm1$ correspond,
respectively, to the $\kappa$ logarithm  and to the q logarithm;
the remaining cases are different. Among these additional
logarithms, and corresponding exponentials, only the cases
$w=\pm1/3$ and $\pm1/2$ satisfy all the requirements discussed in
Sec.~\ref{sec:deformedlogarithm}.

We consider explicitly the case $r= \pm|\kappa|/3 $. Introducing
the parameter $\gamma=\pm2\,|\kappa|/3$, Eq. (\ref{log}) defines a
 generalized logarithm
\begin{eqnarray}
\log_{_{\gamma}}\,(x)=\frac{x^{2\,\gamma}-x^{-\gamma}}{3\,\gamma}
\ ,\label{alog}
\end{eqnarray}
which reduces to the standard ones in the $\gamma\rightarrow0$
limit. This logarithm is an analytical, concave, and increasing
function  for all $x\geq0$,  when $-1/2 <\gamma <1/2$.

If $\gamma$ is positive the asymptotic behaviors for
$x\rightarrow0$ and $x\rightarrow\infty$ are
\begin{eqnarray}
\ln_{_{\gamma}}(x){\atop\stackrel{\textstyle\sim}{\scriptstyle
x\rightarrow {0^+}}} -\frac{x^{-\gamma}}{3\gamma} \ , \quad
\ln_{_{\gamma}}(x){\atop\stackrel{\textstyle\sim}{\scriptstyle
x\rightarrow {+\infty}}}\frac{x^{2\gamma}}{3\gamma} \ . \ \ \
\label{asi}
\end{eqnarray}
Since the logarithm (\ref{alog}) satisfies the duality relation
$\ln_{_{\gamma}}\,(x)=-\ln_{_{-\gamma}}\,\left(1/x\right) \
,\label{condition3}$ the  asymptotic behaviors for $x\rightarrow
0^+$ and $x\rightarrow\infty$ in Eq. (\ref{asi}) are exchanged
when $\gamma<0$.

The corresponding  $\gamma$ exponential is
\begin{eqnarray}
\exp_{_{\gamma}}\,(x) &=& \Bigg[\left(
\frac{1+\sqrt{1-4\,\gamma^3\,x^3}}{2} \right)^{1/3} \nonumber
\\ &+& \left( \frac{1-\sqrt{1-4\,\gamma^3\,x^3}}{2}
\right)^{1/3}\Bigg]^{1/\gamma} \ , \ \ \ \label{aexp}
\end{eqnarray}
which is an analytic, monotonic, and convex function for all $x\in
I\!\!R$ when $-1/2<\gamma<1/2$, and reduces to the standard
exponential in the limit $\gamma\rightarrow0$. The asymptotic
power-law behaviors for $\gamma$ positive are
\begin{equation}
\exp_{_\gamma}(x){\atop\stackrel{\textstyle\sim}{\scriptstyle
x\rightarrow {-\infty}}} \left(3\,\gamma\,|x|\right)^{-1/\gamma
}, \ \mbox{ }
\exp_{_\gamma}(x){\atop\stackrel{\textstyle\sim}{\scriptstyle
x\rightarrow {+\infty}}} \left(3\gamma x\right)^{1/2\,\gamma} \ ,
\end{equation}
where it is clear that the asymptotic behaviors for $x\to +\infty$
and $x\to -\infty$ are exchanged when $\gamma$ changes sign,
coherently with the property $\exp_{\gamma}(x) \exp_{-\gamma}(-x)
= 1$.  Finally the deformed sum given by Eq. (\ref{sd1}) becomes
\begin{eqnarray}
\nonumber
&&\!\!\!\!\!\!\!\!x\oplus\!\!\!\!^{^{^{\scriptscriptstyle
\gamma}}}\,\, y= -3\,\gamma\,x\,y  \nonumber \\ \nonumber
&&\!\!\!\!\!\!\!\!+ x\, \left[ \left(\frac{1+\sqrt{1-4\gamma^3
y^3}}{2} \right)^{1/3}\!\!\!+ \left(\frac{1-\sqrt{1-4\gamma^3
y^3}}{2} \right)^{1/3} \right]^2\\&&\!\!\!\!\!\!\!\!+y\,\left[
\left(\frac{1+\sqrt{1-4\gamma^3 x^3}}{2} \right)^{1/3}\!\!\!+
\left(\frac{1-\sqrt{1-4\gamma^3 x^3}}{2} \right)^{1/3}
\right]^2\nonumber \\ \!\!\!\!\!\!\!\!\label{asum}
\end{eqnarray}
and reduces to the ordinary sum for $\gamma\rightarrow0$.


\sect{Entropies and distributions} \label{sec:entropy}

Having obtained the deformed logarithm as a solution of the
differential equation~(\ref{condif}), the corresponding
generalized entropy follows from Eq. (\ref{defentropy}):
\begin{equation}
S_{_{\kappa,\,r}}(p)=-\sum_{i=1}^Np_{_i}\,\ln_{_{\{{\scriptstyle
\kappa,\,r}\}}}(p_{_i}) =-\sum_{i=1}^Np_{_i}^{1+r}\,
{\displaystyle\frac{p_{_i}^\kappa-p_{_i}^{-\kappa}}{2\,\kappa}} .
\label{entropy1}
\end{equation}
We observe that this class of entropies coincides with the one
introduced by Mittal \cite{Mittal} and Sharma and Taneja
\cite{Taneja1} (MST) and successively derived by Borges and Roditi
in Ref.~\cite{Borges1} by using an approach based on a
biparametric generalization of the Jackson derivative.

Since the entropy is defined in terms of the deformed logarithm
(\ref{log}), the properties (\ref{pd1})-(\ref{pd51}) of
$\ln_{_{\{{\scriptstyle \kappa,\,r}\}}}(x)$ assure that the
entropy (\ref{entropy1}) satisfies many of the properties
satisfied by the standard BGS entropy (\ref{BGSentropy}). In
particular, it is (a) positive definite,
$S_{_{\kappa,\,r}}(p)\geq0$ for $p\in[0,\,1]$; (b) continuous; (c)
symmetric, $S_{_{\kappa,\,r}}(p)=S_{{\kappa,\,r}}(q)$ with
$p\equiv(p_{1},\,\ldots,\,p_{_N})$ and
$q\equiv(p_{{\tau(1)}},\,\ldots,\,p_{{\tau(N)}})$ where $\tau$ is
any permutation from $1$ to $N$; (d) expansible, which means that
$S_{_{\kappa,\,r}}(p)=S_{_{\kappa,\,r}}(q)$ for
$p\equiv(p_{1},\,\ldots,\,p_{N})$ and
$q\equiv(p_{1},\,\ldots,\,p_{_N},\,0,\,\ldots,\,0)$; (e) decisive,
in the sense that $S_{{\kappa,\,r}}(p^{(0)})=0$ where
$p^{(0)}\equiv(0,\,\ldots,\,1,\,\ldots,\,0)$ is a completely
ordered state; (f) maximal, which means that entropy  reach its
maximal value when the distribution is uniform: $Max[S(p)]$ for
$p=p^{(U)}$, with $p^{({\rm U})}\equiv(1/N,\,\ldots,\,1/N)$; and,
finally, (g) concave. Moreover, it will be shown in the next
section that the whole family of entropies (\ref{entropy1})
satisfies the Lesche inequality.

We observe that for a uniform distribution $p^{(U)}$, the entropy
(\ref{entropy1}) assumes the expression
\begin{eqnarray}
S_{_{\kappa,\,r}}\left(p^{(U)}\right)
=-\ln_{_{\{\kappa,r\}}}\left(\frac{1}{N}\right) \ ,
\end{eqnarray}
and only for the case $r=0$, according to Eq. (\ref{pd6}), does it
become
\begin{eqnarray}
S_{_{\kappa}}\left(p^{(U)}\right)=\ln_{_{\{\kappa\}}}(N) \
,\label{Boltzmann}
\end{eqnarray}
which is the generalization of the well-known Boltzmann formula
and  gives  the entropy of a nonextensive microcanonical system as
the deformed logarithm of the number of accessible states of the
system. If the alternative form
\begin{eqnarray}
S_{_{\kappa,\,r}}(p)=\sum_i
p_{_i}\,\ln_{{\{\kappa,\,r\}}}\left(\frac{1}{p_{i}}\right) \
\label{ent3}
\end{eqnarray}
is adopted, the entropy of  a uniform
distribution reduces to Eq. (\ref{Boltzmann}) for any values of
$r$ and $\kappa$.

From a mathematical point of view, the properties of the entropy
(\ref{entropy1}) follow from those of $\ln_{_{\{\kappa,\,r\}}}(x)$
in the range $x\in[0,\,1]$, while the properties of the entropy
(\ref{ent3}) follow from those of $\ln_{_{\{\kappa,\,r\}}}(x)$ in
the range $x\in[1,\,+\infty)$. This justifies our study of the
properties of $\ln_{_{\{\kappa,\,r\}}}(x)$ in the whole range
$x\in[0,\,+\infty)$.

Regarding the relationship  between the entropy of a system and
the entropies of its subsystems, additivity and extensivity
do not hold, in general. However,
it is possible to show that any entropy
belonging to the family (\ref{entropy1}) satisfies an extended
version of the additive and extensive property \cite{Kaniadakis2}.

In fact Eq. (\ref{entropy1}) can be written as
\begin{eqnarray}
S_{_{\kappa,\,r}}(p)=-\langle\ln_{_{\{\kappa,\,r\}}} (p)\rangle \
,\label{ent2}
\end{eqnarray}
which expresses the entropy $S_{_{\kappa,\,r}}(p)$ as the mean
value of $\ln_{_{\{\kappa,\,r\}}}(p)$. Given two systems $A$ and
$B$, with probability distributions $p^A_{_i}$ and $p^B_{_i}$, we
can define a joint system $A\cup B$ with distribution
$p_{_{ij}}^{A\cup B} =p_{_i}^A
\otimes\mbox{\raisebox{-2mm}{\hspace{-4.5mm}$\scriptscriptstyle
{\kappa,r}$}} \hspace{1mm} p_{_j}^B$, where the deformed product $
\otimes\mbox{\raisebox{-2mm}{\hspace{-3.5mm}$\scriptscriptstyle
{\kappa,r}$}} \hspace{1mm} $ is discussed in
Sec.~\ref{sec:algebra}. From Eqs. (\ref{ent2}) and (\ref{sumlog})
it follows
\begin{eqnarray}
S_{_{\kappa,\,r}}(A\cup
B)=S_{_{\kappa,\,r}}(A)+S_{_{\kappa,\,r}}(B) \ .
\end{eqnarray}

\begin{figure*}[ht]
\includegraphics[angle=0,width=0.9\textwidth]{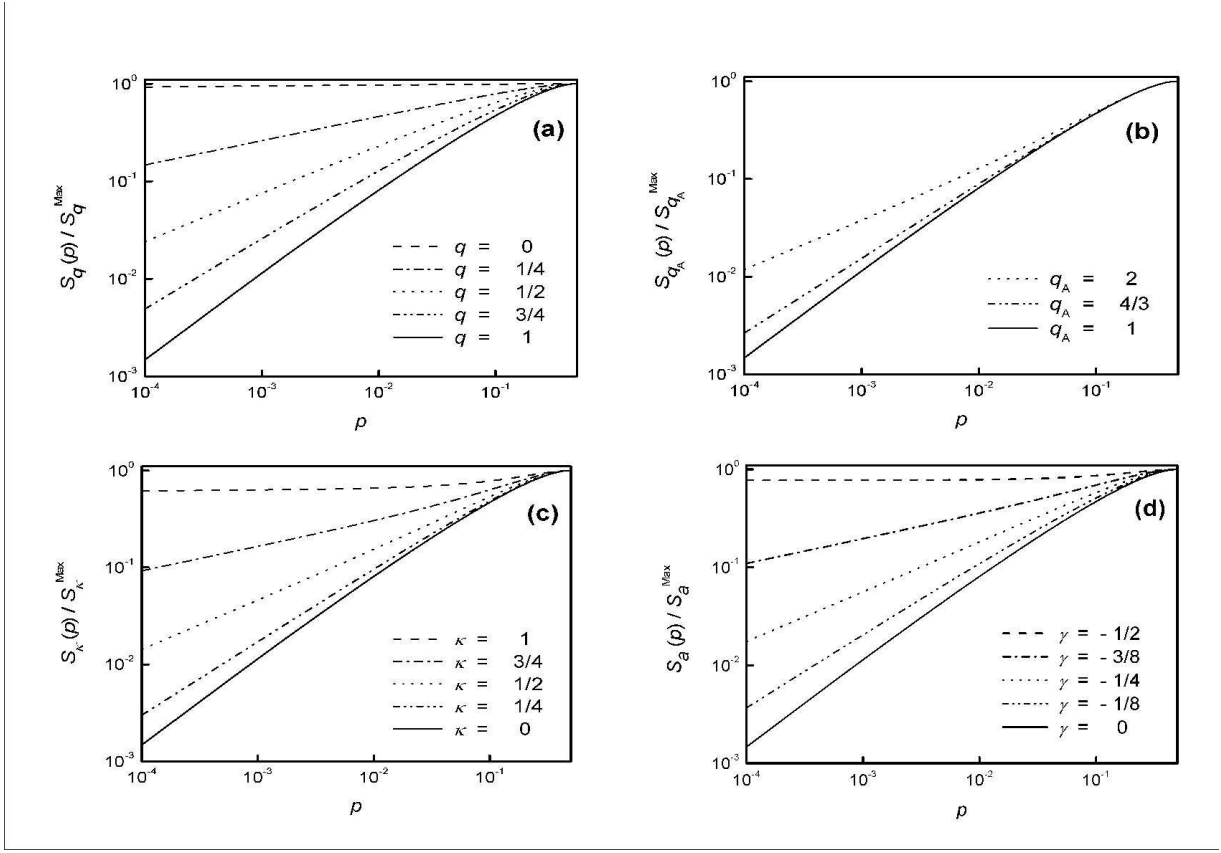}
\caption{Four one-parameter entropies for several values of the
deformed parameter as a function of $p$ in a two-level system:
$({\bfm a})$ Tsallis entropy Eq. (\ref{tsallisent}); $({\bfm b})$
Abe entropy Eq. (\ref{abeent}); $({\bfm c})$ $\kappa$ entropy Eq.
(\ref{kent}); and $({\bfm d})$ $\gamma$ entropy Eq.
(\ref{gammaent}). Broken curves with the same style show entropies
whose corresponding distributions have the same power-law
asymptotic decay $x^{-\nu}$, $\nu = 1$, 4/3, 2, and 4  from top to
bottom; the solid curves show the Shannon entropy.
\label{fig:entropies} } \vspace{10mm}
\end{figure*}

In Fig.~\ref{fig:entropies} we plot four one-parameter entropies
belonging to the family of the MST entropy as functions of $p$ for
system with two states of probabilities $p$ and $1-p$.
(a) is the Tsallis entropy
\begin{equation}
S_{_q}(p)=\sum_{i=1}^N \frac{p_{_i}^q-p_{_i}}{1-q} \
.\label{tsallisent}
\end{equation}
Notice that the entropy (\ref{tsallisent}) is expressed in terms
of the Tsallis logarithm as $S_{_q}(p)=\sum_i
p_{i}\,\ln_{q}(1/p_{i})$, which is different from our choice
(\ref{defentropy}). The property
$\ln_{_{q}}\,(x)=-\ln_{_{2-q}}\,(1/x)$ shows that our choice
$S_{q}(p)= -\sum_i p_{i}\,\ln_{_q}(p_{i})
 = \sum_i p_{i}\,\ln_{2-q}(1/p_{i})$ corresponds to a different
labeling of the entropy  $q\rightarrow2-q$.

(b) is the Abe entropy
\begin{equation}
S_{_{q_{_{\rm A}}}}(p)=-\sum_{i=1}^N \frac{p_{_i}^{q_{_{\rm
A}}}-p_{_i}^{(q_{_{\rm A}}^{-1})}}{q_{_{\rm A}}-q_{_{\rm A}}^{-1}}
\ .\label{abeent}
\end{equation}

(c) is the $\kappa$ entropy
\begin{equation}
S_{_\kappa}(p)=-\sum_{i=1}^N
\frac{p_{_i}^{1+\kappa}-p_{_i}^{1-\kappa}}{2\,\kappa} \
.\label{kent}
\end{equation}
Notice that the entropy (\ref{kent}) due to the property
$\ln_{_{\{{\scriptstyle \kappa}\}}}(1/x)=-\ln_{_{\{{\scriptstyle
\kappa}\}}}(x)$ can be written in the form
$S_{_{\kappa}}(p)=\sum_i p_{i}\, \ln_{_{\{{\scriptstyle
\kappa}\}}}(1/p_{i})=-\sum_i p_{i}\, \ln_{_{\{{\scriptstyle
\kappa}\}}}(p_{i})$ like the Boltzmann-Shannon entropy.

(d) is the $\gamma$ entropy
\begin{equation}
S_{_\gamma}(p)=-\sum_{i=1}^N\frac{p_{_i}^{1+2\,\gamma}-p_{_i}^{1-\gamma}}{3\,\gamma}
\ .\label{gammaent}
\end{equation}
Entropies with the same broken-curve style yield distributions
with the same-power asymptotic behavior $1/x^{\nu}$: $\nu= 1$,
4/3, 2, and 4 from top to bottom; the solid curve shows the
Shannon entropy.

The distribution that optimizes the entropy (\ref{entropy1}) with
the constraints of the canonical ensembles (\ref{con}) is, by
construction,
\begin{eqnarray}
p_{_i}=\alpha\,\exp_{_{\{{\scriptstyle
\kappa,\,r}\}}}\left(-\frac{\beta}{\lambda}\left(E_{_i}-\mu\right)\right)
\ ,\label{dist}
\end{eqnarray}
where we recall that the deformed exponential
$\exp_{_{\{{\scriptstyle \kappa,\,r}\}}}(x)$ is defined as the
inverse of the deformed logarithm, which exists since
$\ln_{_{\{{\scriptstyle \kappa,\,r}\}}}(x)$ is a monotonic
function. The parameter $\mu$ is determined by $\sum _ip_{_i}=1$.

In the $(\kappa,\,r)\to(0,\,0)$ limit, $\lambda=1$ and
$\alpha=e^{-1}$, and Eq. (\ref{dist}) reduces to the well-known
Gibbs distribution
\begin{equation}
p_{_i}=Z(\beta)^{-1}\,\exp(-\beta\,E) \ ,\label{Gibbs}
\end{equation}
where the partition function is given by
$Z(\beta)=\exp(1-\beta\,\mu)=\sum_i\exp(-\beta\,E_{_i})$. The
quantity $\exp(-\beta\,E)$ is named the  Boltzmann factor.
Analogously we can call $\exp_{_{\{{\scriptstyle
\kappa,\,r}\}}}\left(-\beta\,E/\lambda\right)$ the generalized
Boltzmann factor.

We observe that the distribution (\ref{dist}) can not be
factorized as in Eq. (\ref{Gibbs}): the normalization constraint
is fulfilled by fixing $\mu$ in the generalized Boltzmann factor.

\begin{figure*}[ht]
\includegraphics[angle=0,width=0.9\textwidth]{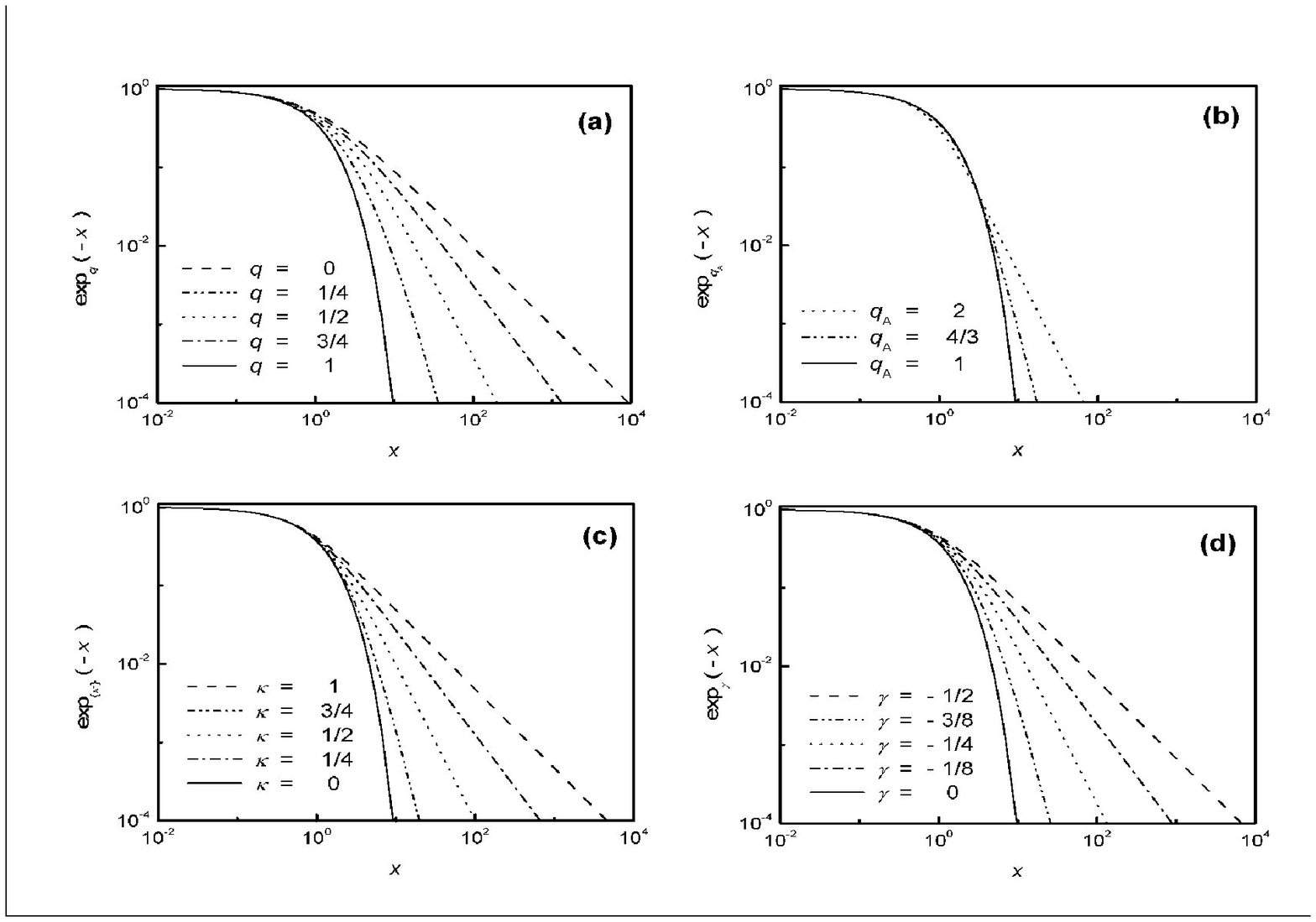}
\caption{The generalized Boltzmann factors that correspond to
entropies in Fig.~\ref{fig:entropies}. \label{fig:distributions} }
\vspace{10mm}
\end{figure*}

Figure~\ref{fig:distributions} shows the generalized Boltzmann
factors corresponding to the four one-parameter entropies of
Fig.~\ref{fig:entropies}. Curves with the same style have the same
asymptotic behavior. Given this constraint and the normalization,
the main difference between the distributions is in the middle
region which joins the linear region ($x\ll1$) and the Zip-Pareto
region ($x\gg1$).
\sect{Lesche inequality} \label{sec:lesche} An important issue is
whether the entropies of the family under consideration are stable
under small changes of the
distribution~\cite{Lesche,Scarfone,Naudts3,Scarfone1}: we want to
demonstrate that, if the two distributions are sufficiently close,
the corresponding relative difference of entropies can be made as
small as one wishes. To this end, we rewrite the entropy
(\ref{defentropy}):
\begin{eqnarray}
\nonumber S(p) \!\!&=&\!\! -\sum_{i=1}^N
p_{_i}\,\Lambda(p_{_i})\nonumber \\
&=&-\sum_{i=1}^N \int\limits_0\limits^{p_{_i}} \frac{d}{dx}\Big[
x\,\Lambda(x)\Big]\,dx \nonumber \\
&=&-\lambda\,\sum_{i=1}^N\int\limits_0^{p_{_i}}
\Lambda\left(\frac{x}{\alpha}\right)\,dx \nonumber \\
&=&-\lambda\,\sum_{i=1}^Nx\,\Lambda\left(\frac{x}{\alpha}\right)
\Bigg|_0^{p_{_i}} +\lambda \sum_{i=1}^N
\int\limits_0\limits^{p_{_i}}
x\,\frac{d}{dx}\left[\Lambda\left(\frac{x}{\alpha}\right)\right]\,dx
\nonumber  \\
&=& -\lambda \sum_{i=1}^N
p_{_i}\,\Lambda\left(\frac{p_{_i}}{\alpha}\right) \!-\!
\sum_{i=1}^N
\!\!\!\int\limits_{-\lambda\,\Lambda(0^+)}\limits^{-\lambda\,\Lambda(p_{_i}/\alpha)}
\!\!\!\!\!\!\!\!\alpha\Lambda^{\!-1\,}\!\!\left(\!-\frac{s}{\lambda}\right)\!ds
, \ \ \ \ \ \ \
\end{eqnarray}
where in the second equality we have used Eq.~(\ref{condif}), in
the last equality we have made the change of variables
$s=-\lambda\,\Lambda(x/\alpha)$, and $\Lambda(0^+)\equiv
\lim_{x\to 0^+} \Lambda(x)$. For the moment we use the notation
$\Lambda(x)$ and ${\cal E}(x)$, since we do not need the specific
form of the deformed logarithm and exponential. Using the fact
that $\Lambda^{-1}(x)={\cal E}(x)$ for the class of entropies
under scrutiny, one finds
\begin{eqnarray}
\nonumber S(p) &=&-\sum_{i=1}^N
\int\limits^{-\lambda\,\Lambda(0^+)}\limits_{-\lambda\,\Lambda(p_{_i}/\alpha)}
\left[p_{_i}-\alpha\,{\cal
E}\left(-\frac{s}{\lambda}\right)\right]\,ds
-\lambda\,\Lambda(0^+)\\
\nonumber &=&
-\sum_{i=1}^N\int\limits^{-\lambda\,\Lambda(0^+)}\limits_{-1}\,\,\,\,
\left[p_{_i}-\alpha\,{\cal E}\left(-\frac{s}{\lambda}\right)
\right]_+\,ds-\lambda\,\Lambda(0^+)\\&=&
\int\limits_{-1}\limits^{-\lambda\,\Lambda(0^+)}
\left[1-A(p,\,s)\right]\,ds-1 \ ,\label{ent1} \
\end{eqnarray}
where in the second equality it was used  $\sum_i p_{_i}=1$, that
$\alpha\,{\cal E}(1/\lambda)\geq\alpha\, {\cal
E}(-s/\lambda)>p_{_i} $ for $-1\leq s<-\lambda\,
\Lambda(p_{_i}/\alpha)$, and the definitions $[x]_+ \equiv
\max(x,0)$ and
\begin{eqnarray}
A(p,\,s) \equiv \sum_{i=1}^N\left[p_{_i}- \alpha\,{\cal
E}\left(-\frac{s}{\lambda}\right)\right]_+ \ .\label{a}
\end{eqnarray}
From now on we revert to the notation $\exp_{_{\{\kappa,r\}}}(x)$
and  $\ln_{_{\{\kappa,r\}}}(x)$. We remark that the upper limit of
the integral $s_{\text{m}}\equiv -\lambda\,\Lambda(0^+)$ in
Eq.~(\ref{ent1}) is $s_{\text{m}}=+\infty$ for $r\not=|\kappa|$
and $s_{\text{m}} =1/|2\kappa|$ for $r =|\kappa|$; see
Eq.~(\ref{pd7}).

The definition of  $A(p,\,s)$, Eq.~(\ref{a}), implies that~\cite{Abe4}:
\begin{eqnarray}
\Big|A(p,\,s)-A(q,\,s)\Big|\leq
\sum_{i=1}^N|p_{_i}-q_{_i}|
\equiv
|\!|p-q|\!|_1
\ , \label{pp1}
\end{eqnarray}
and, for values of  $s\geq-\lambda\,\ln_{_{\{\kappa,r\}}}(1/N)$,
\begin{eqnarray}
\!1\!-N\alpha
\,\exp_{_{\{\kappa,r\}}}\!\left(-\frac{s}{\lambda}\right)\!\!&=&\!\!
\left[ \sum_{i=1}^N\left(p_{_i}-
\alpha\,\exp_{_{\{\kappa,r\}}}\!\!\left(-\frac{s}{\lambda}\right)\right)
\!\right]_+ \nonumber \\ &<&\sum_{i=1}^{N} p_{_i}=1 \ ,
\end{eqnarray}
from which it follows that
\begin{eqnarray}
\Big|A(p,\,s)-A(q,\,s)\Big|<N\,
\alpha\,\exp_{_{\{\kappa,r\}}}\left(-\frac{s}{\lambda}\right)
 \ .\label{pp2}
\end{eqnarray}

From Eq. (\ref{ent1}) the absolute difference of the entropies of
two different distributions
$p\equiv\{p_{_i}\}_{_{i=1,\,\ldots,\,N}}$ and
$q\equiv\{q_{_i}\}_{_{i=1,\,\ldots,\,N}}$ satisfies
\begin{eqnarray}
\nonumber \!\!\!\!
\Big|S_{_{\kappa,\,r}}(p)-S_{_{\kappa,\,r}}(q)\Big|&=&
\Big|\int\limits_{-1}\limits^{s_{_{\rm m}}}
\left[A(p,\,s)-A(q,\,s)\right]\,ds\Big| \nonumber \\
&\leq&\int\limits_{-1}\limits^{s_{_{\rm m}}}\Big|
A(p,\,s)-A(q,\,s)\Big|\,ds \nonumber \\
&=&\int\limits_{-1}\limits^\ell\Big| A(p,\,s)-A(q,\,s)\Big|\,ds
\nonumber \\ &+& \int\limits_\ell\limits^{s_{_{\rm m}}}\Big|
A(p,\,s)- A(q,\,s)\Big|\,ds \ . \ \ \ \ \ \ \label{eq1}
\end{eqnarray}
Choosing $ -\lambda\,\ln_{_{\{\kappa,r\}}}(1/N)  \leq \ell
<s_{\text{m}}$, by using Eq (\ref{pp1}) in the first integral and
Eq. (\ref{pp2}) in the second integral of Eq. (\ref{eq1}), we
obtain
\begin{eqnarray}
\Big|S_{_{\kappa,\,r}}(p)-S_{_{\kappa,\,r}}(q)\Big| &\leq&
|\!|p-q|\!|_1\,\left(\ell+1\right)\nonumber \\
&+&N\,\alpha\int\limits_\ell\limits^{s_{_{\rm m}}}
\exp_{_{\{\kappa,r\}}}\!\!\left(-\frac{s}{\lambda}\right)\,ds \ .
\ \ \ \ \ \  \label{eee1}
\end{eqnarray}
In particular  Eq.~(\ref{eee1}) holds for that value $\bar{\ell}$
that minimizes the right-hand side of Eq. (\ref{eee1}),
\begin{eqnarray}
\bar{\ell}
=-\lambda\,\ln_{_{\{\kappa,r\}}}\!\!\left(\frac{|\!|p-q|\!|_1}{\alpha\,N}\right)
\ ,\label{eee}
\end{eqnarray}
as long as
$\bar{\ell}\geq-\lambda\,\ln_{_{\{\kappa,r\}}}(1/N)$,
which is true when
\begin{eqnarray}
|\!|p-q|\!|_1\leq\alpha \ ,\label{eqq}
\end{eqnarray}
 i.e., for sufficiently close distributions, according to the
metric $|\!|\ldots|\!|_1$. Introducing Eqs.~(\ref{eqq}) and
(\ref{eee}) in Eq. (\ref{eee1}) and performing the integration
using the result~(\ref{integralcalE}), we obtain
\begin{equation}
\label{deltaSdeltap}
\Big|S_{_{\kappa,\,r}}(p)-S_{_{\kappa,\,r}}(q)\Big|\leq
|\!|p-q|\!|_1\,\left[1- \ln_{_{\{\kappa,r\}}}\!\!
\left(\frac{|\!|p-q|\!|_1}{N}\right)\right] \ ,
\end{equation}
and the relative difference of entropies can be written as
\begin{equation}
\Bigg|\frac{S_{_{\kappa,\,r}}(p)-S_{_{\kappa,\,r}}(q)}{S_{\rm
max}}\Bigg|\leq F_{_{\kappa,r}}(|\!|p-q|\!|_1,\,N) \ ,
\end{equation}
with
\begin{equation}
F_{_{\kappa,r}}(|\!|p-q|\!|_1,\,N)=\frac{|\!|p-q|\!|_1}{\ln_{_{\{\kappa,\,-r\}}}(N)}\,\left[1-
\ln_{_{\{\kappa,r\}}}\!\!\left(\frac{|\!|p-q|\!|_1}{N}\right)\right]
\ ,
\end{equation}
because $S_{\rm max}\equiv\ln_{_{\{\kappa,\,-r\}}}(N)$.

This result demonstrates that if the two distributions are
sufficiently close the corresponding absolute difference of
entropies can be made as small as one wishes, since
Eq.~(\ref{pd5}) implies that $\lim_{x\to 0^+}
x\,\ln_{_{\{\kappa,r\}}}(x) =0$.

In particular, the Lesche inequality for the family of entropies under
scrutiny is valid also
in the thermodynamic limit $N\to\infty$
\begin{equation}
\lim_{|\!|p-p^\prime|\!|\to
0^+}\,\lim_{N\to\infty}F_{_{\kappa,r}}(|\!|p-q|\!|_1,\,N)=0 \ .
\end{equation}
This last result is not trivial, since the thermodynamical limit
introduces nonanalytical behaviors that could produce finite
entropy differences between probability distributions
infinitesimally close. We conclude this section by noting that
Lesche stability of the $(\kappa,r)$ family of entropies follows
also from the general proof given in \cite{Naudts3}.


\sect{Conclusions}
\label{sec:conclusion}
In order to unify several entropic forms,
the canonical MaxEnt principle has been applied to a generic
trace-form entropy obtaining the
differential-functional equation (\ref{condint}) for the corresponding
generalized logarithm, when the ensuing distribution function
is required to be expressed in terms of the generalized exponential
through the  natural relation  (\ref{distribution}).

The solution of this equation yields the biparametric family of
logarithms
\begin{eqnarray}
\ln_{_{\{\kappa,\,r\}}}(x)=
x^r\,{\displaystyle\frac{x^\kappa-x^{-\kappa}}{2\,\kappa}} \ ;
\end{eqnarray}
the corresponding entropy \cite{Mittal,Taneja1,Borges1} is
\begin{equation}
\label{stentropy} S_{_{\kappa,\,r}}(p)=-\sum_{i=1}^N p_{_i}^{1+r}
\,\frac{p_{_i}^\kappa-p_{_i}^{-\kappa}}{2\,\kappa} \quad .
\label{entroconclu}
\end{equation}
This entropy is a mathematically and physically sound entropy when
the parameters $\kappa$ and $r$ belong to the region shown in
Fig.~\ref{fig:parameterRegion} and, therefore, the $(\kappa,\,r)$
logarithm satisfies the set of properties
(\ref{pd1})-(\ref{pd51}). In particular these entropies satisfy
the Lesche stability condition.

Distribution functions obtained by extremizing the entropy
(\ref{entroconclu}) have power-law asymptotic behaviors: such
behaviors could be relevant for describing anomalous systems; a
comparison between several one-parameter distribution functions is
shown in Fig.~\ref{fig:distributions}.

In addition, we have shown that several important one-parameter
generalized entropies (Tsallis entropy, Abe entropy and $\kappa$
entropy) are specific cases of this family; when the deformation
parameters vanish, the family collapses to the Shannon entropy.

Our approach yielded also new one-parameter logarithms
belonging to this family, whose corresponding exponentials
can be explicitly given by algebraic methods.

There remains the question of the relevance of each mathematically
sound entropy to specific physical situations. In fact a  wide
class of deformed logarithms satisfy a set of reasonable
mathematical properties and physical constraints, in particular
concavity, related to  thermodynamic stability, and the Lesche
inequality, related to the experimental robustness.

\acknowledgments This work was partially supported by MIUR
(Ministero dell'Istruzione, del\-l'Uni\-ver\-si\-t\`a e della
Ricerca) under MIUR-PRIN-2003 project ``Theoretical Physics of the
Nucleus and the Many-Body Systems.''

\end{document}